\documentclass[preprint,aps,prd,superscriptaddress,nofootinbib,tightenlines]{revtex4}
\usepackage[english]{babel}

\usepackage[letterpaper,top=2cm,bottom=2cm,left=3cm,right=3cm,marginparwidth=1.75cm]{geometry}

\usepackage{amsfonts}
\usepackage{multirow}
\usepackage{makecell}
\usepackage{mathrsfs}
\usepackage{graphicx}
\usepackage{amsmath}
\usepackage{amssymb}
\usepackage{bm}
\usepackage{bbm}
\usepackage{color}
\usepackage{ulem}
\usepackage{array}
\usepackage{diagbox}
\usepackage{float}

\newcommand{\nn}{\nonumber}

\newcommand{\beq}{\begin{equation}}
\newcommand{\eeq}{\end{equation}}
\newcommand{\bqa}{\begin{eqnarray}}
\newcommand{\eqa}{\end{eqnarray}}

\newcommand{\bseq}{\begin{subequations}}
\newcommand{\eseq}{\end{subequations}}

\makeatletter

\begin{document}

\title{Next-to-next-to-leading-order QCD corrections to ${}^3S_1^{(8)}$ 
gluon fragmentation function for quarkonium }

\author{Feng Feng~\footnote{f.feng@outlook.com}}
\affiliation{China University of Mining and Technology, Beijing 100083, China\vspace{0.2cm}}
\affiliation{Institute of High Energy Physics, Chinese Academy of Sciences, Beijing 100049, China\vspace{0.2cm}}
\author{Yu Jia~\footnote{yjia@m.scnu.edu.cn}}
\affiliation{State Key Laboratory of Nuclear Physics and Technology, Institute of Quantum Matter,  South China Normal University, Guangzhou 510006, China\vspace{0.2cm}}
\affiliation{Guangdong Basic Research Center of Excellence for Structure and Fundamental Interactions of Matter, Guangdong Provincial Key Laboratory of Nuclear Science, Guangzhou 510006, China\vspace{0.2cm}}
\author{Wen-Long Sang~\footnote{wlsang@swu.edu.cn}}
\affiliation{School of Physical Science and Technology, Southwest University, Chongqing 400700, China
}


\date{\today} 

\begin{abstract}
We present the first computation of the next-to-next-to-leading-order (NNLO) QCD corrections to the ${}^3S_1^{(8)}$ gluon fragmentation function for quarkonium within the nonrelativistic QCD (NRQCD) factorization framework, 
accurate to the lowest order in the velocity expansion. 
The calculation is performed with high numerical precision and encompasses both polarized and unpolarized cases. 
We find that the NNLO corrections are positive and substantial across most of the $z$ region.
Furthermore, the logarithmic singularities near the endpoint
$z\to 1$ are fully reconstructed, providing essential inputs for 
future threshold resummation beyond leading-logarithmic accuracy. 
Combined with threshold-resummed formulas in the large-$z$ 
region, our results yield phenomenologically viable inputs for the
$^3S_1^{(8)}$ gluon fragmentation function. 
This enables a more reliable description of large-$p_T$ $J/\psi$ ($\psi'$) 
and $\chi_{cJ}$ production and polarization at hadron colliders, 
representing a crucial step toward a definitive test of the color-octet mechanism.
\end{abstract}

\maketitle

\section{Introduction} 

Inclusive quarkonium production at high-energy colliders has long served as a 
pivotal testing ground for QCD dynamics~\cite{QuarkoniumWorkingGroup:2004kpm,Brambilla:2010cs}. 
The theoretical landscape was revolutionized by the nonrelativistic QCD (NRQCD) 
factorization framework~\cite{Bodwin:1994jh},  a development propelled 
by two key theoretical breakthroughs in the mid-1990s: 
the {\it color-octet mechanism} proposed by Braaten and Fleming to resolve the $\psi'$ surplus 
puzzle observed at the {\tt Tevatron}~\cite{Braaten:1994vv}, and the prediction by Cho and Wise of 
nearly 100\% transverse polarization for $J/\psi$ at large $p_T$~\cite{Cho:1994ih}.
Unlike the color-singlet model, the NRQCD approach posits that a color-octet spin-triplet 
$c\bar{c}$ pair (e.g., $c\bar{c}({}^3S_1^{(8)})$) produced at short distances 
has a non-negligible probability to hadronize into a $J/\psi$ ($\psi'$) via 
soft gluon radiation without a spin flip~\cite{Bodwin:1994jh}. 
However, subsequent {\tt Tevatron}~\cite{CDF:2000pfk,CDF:2007msx} and 
{\tt LHC}~\cite{ALICE:2011gej,CMS:2013gbz,LHCb:2013izl} measurements have observed unpolarized or 
slightly longitudinally polarized $J/\psi$ at large $p_T$, 
in direct contradiction with the Cho-Wise prediction. 
This alarming discrepancy has stimulated a vast body of research over the past three decades~~\cite{Cho:1995vh,Cho:1995ce,Beneke:1996yw,Petrelli:1997ge,Braaten:1999qk,Kramer:2001hh,Lansberg:2006dh,Campbell:2007ws,
Lansberg:2008gk,Faccioli:2010kd,Baranov:2011ib,Kang:2011mg,Gong:2008sn,Gong:2008hk,Gong:2008ft,Gong:2012ug,Zhang:2014ybe,
Butenschoen:2009zy,Butenschoen:2010rq,Butenschoen:2011yh,Butenschoen:2012px,Butenschoen:2014dra,Ma:2010vd,Ma:2010yw,Ma:2010jj,Chao:2012iv,Shao:2014fca,Shao:2014yta,Han:2014jya,Brambilla:2022ayc,Brambilla:2024iqg}
(For recent reviews on inclusive charmonium hadroproduction, see Refs.~\cite{Lansberg:2019adr,Chen:2021tmf,Chung:2022uih,Andronic:2015wma}).

In the limit $p_T\gg M_H$, the QCD factorization theorem~\cite{Collins:1989gx} dictates 
that the inclusive quarkonium $H$ production is dominated by single-parton 
fragmentation~\cite{Braaten:1993rw}
(Similarly, double-parton fragmentation provides an effective description in the 
intermediate-$p_T$ regime~\cite{Kang:2011zza,Kang:2014tta,Ma:2014svb}).
A key advantage of the fragmentation function approach over fixed-order NRQCD is its ability to 
systematically resum large logarithms of $ \ln(p_T^2/m_Q^2) $ through Dokshitzer-Gribov-Lipatov-Altarelli-Parisi (DGLAP) evolution equations~\cite{Gribov:1972ri,Altarelli:1977zs,Dokshitzer:1977sg}~\footnote{
To date, phenomenological conclusions for quarkonium production have predominantly relied on next-to-leading order 
(NLO) NRQCD predictions~\cite{Gong:2008sn,Gong:2008hk,Gong:2008ft,Gong:2012ug,Zhang:2014ybe,Butenschoen:2009zy,Butenschoen:2010rq,Butenschoen:2011yh,Butenschoen:2012px,Butenschoen:2014dra,Ma:2010vd,Ma:2010yw,Ma:2010jj,Chao:2012iv,Shao:2014fca,Shao:2014yta,Han:2014jya}.
While fixed-order NRQCD has achieved notable successes, it suffers from poor perturbative convergence and 
severe inconsistencies in the fitted color-octet long-distance matrix elements (LDMEs) across different analyses~\cite{Lansberg:2019adr,Chen:2021tmf,Chung:2022uih}.
These shortcomings largely arise because the transverse momentum and heavy quark mass scales are not properly disentangled within the short-distance coefficients. Moreover, extending these fixed-order calculations to next-to-next-to-leading order (NNLO) remains technically prohibitive. }.
At the {\tt LHC}, gluon fragmentation emerges as the dominant production mechanism at large $p_T$.
Phenomenological analyses combining the fragmentation contribution with the NLO NRQCD prediction for hadroproduction of
prompt $J/\psi (\psi^\prime)$ and $\chi_{cJ}$ were performed by Bodwin {\it et al.}~\cite{Bodwin:2014gia,Bodwin:2015iua}.

Fragmentation functions are universal yet nonperturbative quantities that 
generally defy first-principles calculations, such as lattice QCD. 
Consequently, light-hadron fragmentation functions must be extracted from experimental data 
(see, e.g., Refs.~\cite{Bertone:2017tyb,Borsa:2022vvp,AbdulKhalek:2022laj,Gao:2024nkz,Gao:2025hlm} for recent global fits to $\pi$, $K$, and $p/\bar{p}$ fragmentation functions). 
In contrast, quarkonium fragmentation functions are not genuinely nonperturbative. 
Owing to the hierarchy $ m_Q \gg \Lambda_{\rm QCD} $ and the asymptotic freedom of QCD, 
NRQCD factorization~\cite{Bodwin:1994jh} allows the gluon fragmentation function for a quarkonium 
$H$ to be expressed as a sum of perturbatively calculable short-distance coefficients (SDCs) multiplied by 
long-distance matrix elements (LDMEs):
\beq
D_{g \to H}(z,\mu)=\sum_n d_n(z,\mu) \langle 0\vert {\mathcal O^H_n} \vert 0 \rangle,
\label{NRQCD:factorization:frag}
\eeq
where $z$ is the light-cone momentum fraction of $H$ relative to the parent gluon, 
$\mu$ is the renormalization scale, and $n$ denotes the color, spin, and orbital 
quantum numbers of the intermediate $c\bar{c}$ state.
 
The most relevant channels for  $J/\psi (\psi')$ production include 
the states $n={}^3S_1^{(1)}$,  $^3P_J^{(8)}$, $^3S_1^{(8)}$ and $^1S_0^{(8)}$. 
The corresponding SDCs in Eq.~\eqref{NRQCD:factorization:frag} up to ${\mathcal O}(\alpha_s^2)$
have been well established~\cite{Braaten:1993rw,Braaten:1994kd,Ma:1995ci,Beneke:1995yb,Braaten:1996rp,Braaten:2000pc} (see Refs.~\cite{Ma:2013yla,Ma:2015yka} for a summary).  
Furthermore, $\mathcal{O}(\alpha_s^3)$ corrections to these SDCs have been computed 
for all channels except the $^3S_1^{(8)}$ state~\cite{Zhang:2017xoj,Braaten:1995cj,Artoisenet:2018dbs,Feng:2018ulg,Zhang:2018mlo,Zhang:2020atv}. 
Relativistic corrections to the color-octet gluon fragmentation functions for these channels have also been explored~\cite{Bodwin:2003wh,Bodwin:2012xc,He:2026nzm}.

The ${}^3S_1^{(8)}$ gluon fragmentation function takes center stage in
the color-octet mechanism for $J/\psi$ hadroproduction.
At leading order (LO), the SDC is  
proportional to $\alpha_s \delta(1-z)$ and receives contributions solely from
transversely polarized $c\bar{c}({}^3S_1^{(8)})$ pair~\cite{Braaten:1994kd}.
While NLO QCD corrections were initially explored in 
the mid-1990s~\cite{Ma:1995ci,Beneke:1995yb}, a complete NLO calculation was later performed by Braaten and Lee~\cite{Braaten:2000pc}. Their result agrees with Ref.~\cite{Beneke:1995yb} for the longitudinally polarized 
${\cal O}(\alpha_s^2)$ SDC but differs significantly from Ref.~\cite{Ma:1995ci} for the transversely polarized counterpart. 
By identifying a minor error in Ref.~\cite{Braaten:2000pc}, Ma, Qiu, and Zhang 
finally established the correct ${\cal O}(\alpha_s^2)$ SDCs 
for this channel~\cite{Ma:2013yla,Ma:2015yka}.

To achieve a reliable description for $J/\psi$ production and polarization 
at large transverse momentum,  
it is imperative to compute the $^3S_1^{(8)}$ gluon fragmentation function 
to ${\cal O}(\alpha^3_s)$ accuracy, matching the level achieved for 
the ${}^3S_1^{(1)}$,  $^3P_J^{(8)}$, and $^1S_0^{(8)}$ channels. 
The primary goal of this work is to fill this long-standing gap  
by accomplishing the NNLO calculation 
for the ${}^3S_1^{(8)}$ gluon fragmentation function in 
both transversely and longitudinally polarized cases. 
Our results indicate that the NNLO QCD corrections are significant. 
Although we present the results with very high numerical precision, 
we also provide parameterized forms for the polarized SDCs that are sufficiently accurate 
for phenomenological applications.  Additionally, we reconstruct the asymptotic behaviors around the upper $z=1$,
which are useful for performing threshold resummation beyond leading-logarithmic accuracy. 
By incorporating our NNLO predictions with the threshold-resummed expressions valid at large $z$~\cite{Chung:2024jfk,Chung:2026mii}, we establish a phenomenologically viable input for the ${}^3S_1^{(8)}$ gluon fragmentation function. This synthesis significantly improves the theoretical precision for predicting the production rates 
and polarization of large-$p_T$ $J/\psi$ ($\psi'$) and $\chi_{cJ}$ at hadron colliders, thereby facilitating a 
rigorous test of the color-octet mechanism.

\section{NRQCD factorization to ${}^3S_1^{(8)}$ gluon fragmentation function for quarkonium}

We begin with the Collins-Soper definition of the gluon fragmentation function for a specific hadron 
$H$~\cite{Collins:1981uw} (see also Ref.~\cite{Bodwin:2012xc}):
\bqa
\label{CS:def:Fragmentation:Function}
D_{g \to H}(z,\mu) &=&
\frac{-g_{\mu \nu}z^{d-3} }{ 2\pi k^+ (N_c^2-1)(d-2) }
\int_{-\infty}^{+\infty} \!dx^- \, e^{-i k^+ x^-} \langle 0 | G^{+\mu}_c(0)
\Phi^\dagger(0,0,{\bf 0}_\perp)_{cb}
\nn\\
&\times & {\tt P}_{H(P)}
\;\Phi(0,x^-,{\bf 0}_\perp)_{ba} \; G^{+\nu}_a(0,x^-,{\bf 0}_\perp) \vert 0 \rangle,
\eqa
where $G^a_{\mu\nu}$ is the gluon field strength,
$k$ is the momentum injected by the gluon field strength operator, 
and $d=4-2\epsilon$ denotes the space-time dimensions. 
The light-like gauge link $\Phi$ is inserted to ensure
the gauge invariance of the fragmentation function.

The out-state projector ${\tt P}_{H(P)}$ in Eq.~\eqref{CS:def:Fragmentation:Function} is defined as
\beq
{\tt P}_{H(P)} = \sum_{X} \vert H(P)+X\rangle \langle H(P)+X\vert,
\label{Projection:H:out:state}
\eeq
which projects onto a state containing the hadron $H$ 
with four-momentum $P^\mu=(zk^+, M_H^2/(2zk^+),{\bf 0}_\perp)$
along with unobserved hadrons. The polarization of 
$H$ is implicitly summed over in Eq.~\eqref{Projection:H:out:state}.

Hereafter, we consider $H$ to be a quarkonium state, such as $J/\psi$ or $\chi_{cJ}$.
We focus specifically on the contribution from a color-octet spin-triplet 
$c\bar{c}$ pair that hadronizes into $H$. 
According to the NRQCD factorization formula~\cite{Bodwin:1994jh}, the 
${}^3S_1^{(8)}$ channel contributes to the gluon-to-$H$ 
fragmentation function as
\beq
D_\lambda(z,\mu) =  \frac{d_\lambda(z,\mu)}{m^3}    \times 
{\langle 0\vert  {\mathcal O}^H({}^3S_1^{(8)}) \vert 0 \rangle \over (d-1)(N_c^2-1)} + \cdots,
\eeq
where $\lambda=T, L$ designates the transversely and longitudinally polarized $c\bar{c}$ pair, respectively.
The color-octet spin-triplet NRQCD production operator is defined as~\cite{Bodwin:1994jh}~\footnote{It was pointed out by 
Nayak, Qiu, and Sterman~\cite{Nayak:2005rt,Nayak:2006fm} that the original NRQCD color-octet production operator~\cite{Bodwin:1994jh} lacks gauge invariance and must be supplemented with extra gauge links. 
Because the inclusion of these links does not alter the NNLO SDCs at lowest order in 
$v$, we omit this complication here.} 
\beq
{\mathcal O}^H({}^3S_1^{(8)}) = \chi^\dagger\sigma^i T^a \psi \, {\tt P}_{H(P)} \,
\psi^\dagger\sigma^i T^a \chi.
\eeq
Note that the projector ${\tt P}_{H(P)}$ is defined
in the rest frame of the $c\bar{c}$ pair.

The coefficient functions $d_{\lambda}(z,\mu)$ admit a perturbative expansion in powers of  $\alpha_s$:
\beq
d_{\lambda}(z,\mu) = \alpha_s \, d_{\lambda}^{(0)}(z) + \alpha_s^2 \, d_{\lambda}^{(1)}(z,\mu) + \alpha_s^3 \, d_{\lambda}^{(2)}(z,\mu) + \cdots
\eeq

The LO SDCs take a simple form~\cite{Braaten:1994kd}: 
\beq
d_{T}^{(0)}(z) = \pi\delta(1-z), \qquad d_L^{(0)}(z)=0,\qquad d^{(0)}(z) = d_T^{(0)}(z) + d_L^{(0)}(z).
\eeq

The NLO QCD corrections to $d_\lambda(z,\mu)$ are given by~\cite{Ma:2013yla,Ma:2015yka}: 
\bseq
\bqa
d_T^{(1)}(z,\mu) &=&  \frac{3(z^2-z+1)^2}{z(1-z)_+}\ln\frac{\mu^2}{4m^2} -\frac{6(z^2-z+1)^2 }{z} \left[\frac{\ln (1-z)}{(1-z)}\right]_+ 
\\ 
&-& \frac{3 (z^2-z+1)^2}{z(1-z)_+} + A(\mu)\delta(1-z),
\nn\\
d_L^{(1)}(z,\mu) &=&  {3(1-z)\over z},
\\
d^{(1)}(z,\mu) &=& d_T^{(1)}(z,\mu) + d_L^{(1)}(z,\mu),
\eqa
\eseq
where $A(\mu)$ is defined as
\begin{equation}
A(\mu) = \frac{\beta_0}{N_c} \left[ \ln\frac{\mu^2}{4m^2} + 
\frac{13}{3} \right] + \frac{4}{N_c^2} - \frac{\pi^2}{3} + \frac{16}{3}\ln2,
\end{equation}
with $\beta_0 = {11\over 6} N_c-{1\over 3} n_L$
(where $N_c=3$ is the number of colors and $n_L=3$ is the number of light quarks). 
Note that $d^{(1)}_L(z,\mu)$ is independent of $\mu$.

Physically, the NNLO SDCs are expected to take the form:
\beq
d_{\lambda}^{(2)}(z,\mu) = C^{\lambda}_2(z) \ln^2\frac{\mu^2}{m^2} + 
C^{\lambda}_1(z) \ln\frac{\mu^2}{m^2} + \widetilde{C}^{\lambda}_{0}(z) \, n_L + C^{\lambda}_0(z)
\label{NNLO:SDCs:decomposition}
\eeq
where $\lambda=T, L$.
Our primary goal is to determine the coefficient functions in Eq.~\eqref{NNLO:SDCs:decomposition} 
either analytically or with high numerical precision.

\section{Outline of calculation}
\label{sec:outline:calculation}

\begin{figure}[tb]
\centering
\includegraphics[width=0.8\textwidth]{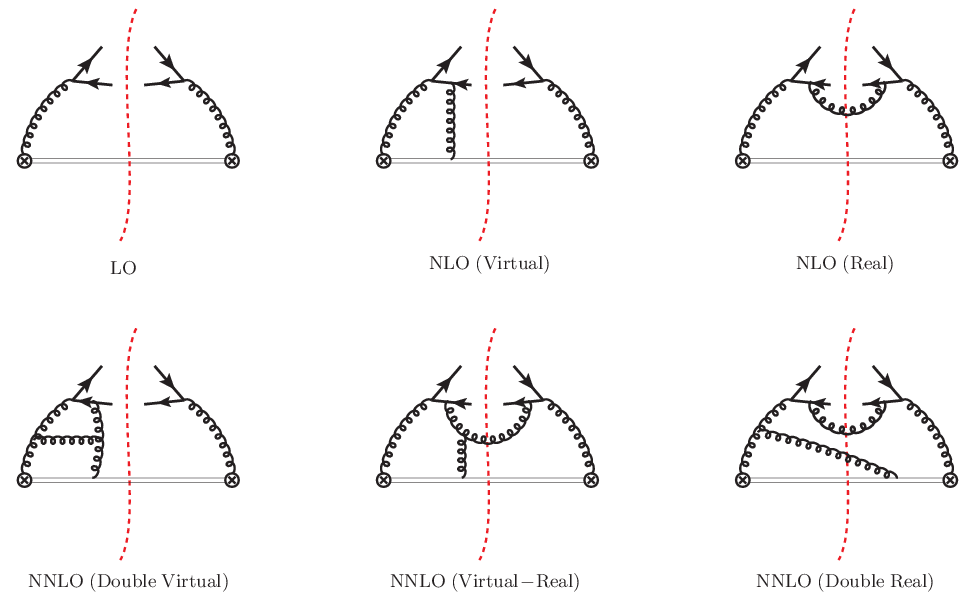}
\caption{Representative Feynman diagrams 
for $g\to c\bar{c}({}^3S_1^{(8)})$ gluon fragmentation function up to NNLO in $\alpha_s$.
The thin double line represents an eikonal line.
\label{Cut:diagram:NNLO}}
\end{figure}

In this section, we outline the NNLO calculation of the ${}^3S_1^{(8)}$ gluon fragmentation
function into a quarkonium. Following the perturbative matching ansatz, 
we replace a physical quarkonium state $H$ with
a free $c\bar{c}({}^3S_1^{(8)},P^\mu)$ pair, where $P^+=k^+/z$. The fragmentation function
in Eq.~\eqref{CS:def:Fragmentation:Function} is then computed order by order in $\alpha_s$. 
Throughout this work, we adopt the Feynman gauge and employ dimensional regularization 
to regulate both UV and IR divergences. The standard QCD Feynman rules, 
along with the eikonal propagator and vertices~\cite{Collins:1981uw}, are implemented in 
\textsf{Qgraf}~\cite{Nogueira:1991ex} and \textsf{FeynArts}~\cite{Hahn:2000kx} to 
generate the Feynman diagrams and their corresponding amplitudes up to NNLO. 
Representative Feynman diagrams for $g\to c\bar{c}({}^3S_1^{(8)})$ 
fragmentation up to NNLO are presented in Fig.~\ref{Cut:diagram:NNLO}.

As shown in Fig.~\ref{Cut:diagram:NNLO}, the NNLO contributions fall into three categories: 
double virtual, virtual-real, and double real.  
The double virtual sector yields terms proportional solely to $\delta(1-z)$. 
Although we have computed the corresponding coefficients with high precision, 
even analytically reconstructing all the poles, they are excluded 
from this work for two reasons.
First, fixed-order NRQCD predictions for $ d_T(z,\mu) $ break down 
near the upper endpoint $z= 1$~\cite{Braaten:2000pc}, because this region is dominated by threshold logarithms 
that must be resummed to yield physically meaningful predictions~\cite{Chung:2024jfk,Chung:2026mii}~\footnote{The 
${}^3S_1^{(8)}$ gluon fragmentation function has also been computed to NLO in $ \alpha_s $ within the 
soft gluon factorization (SGF) framework~\cite{Ma:2017xno}, where the leading threshold logarithms as 
$z\to 1$ have been resummed~\cite{Chen:2021hzo}.}. 
Second, isolating all singular distributions near $z\to 1$ 
in the virtual-real and double real sectors is technically challenging for our numerical method. 
Furthermore, uncancelled infrared poles accompanied by $\delta(1-z)$ are expected to survive. 
These poles must be absorbed into the ${}^3S_1^{(8)}$ NRQCD LDME, whose required two-loop 
anomalous dimensions are, to our knowledge, currently unavailable in the literature. We plan to address the double-virtual contribution in future work.
 
To facilitate the perturbative calculation, we employ the standard covariant projector to project the $c\bar{c}$
pair onto a spin-triplet and color-octet state:
\beq
\Pi_8^a(\lambda) = \frac{1}{2\sqrt{2}}\left(P\!\!\!\!\slash- 2m\right)\,{\epsilon\!\!\!\slash^{(\lambda)\,*}}   \otimes \sqrt{2} T^a.
\label{spin:color:projectors}
\eeq
At the lowest order in the velocity expansion, the relative momentum between the 
$c$ and $\bar{c}$ quarks is neglected ($P^2=4m^2$),  
which effectively restricts the pair to an $S$-wave state. Here,
$\epsilon^{(\lambda)}$ in \eqref{spin:color:projectors} 
in Eq.~\eqref{spin:color:projectors} denotes the polarization vector of the 
$c\bar{c}$ pair pair with helicity $\lambda$. 

Dirac and color traces are evaluated using \textsf{FeynCalc}/\textsf{FormLink}~\cite{Mertig:1990an,Feng:2012tk}, 
and the resulting amplitudes are further simplified via partial fraction decomposition 
with \textsf{Apart}~\cite{Feng:2012iq}.

To automate the formidable NNLO calculation using modern multi-loop tools, 
we apply the {\it reverse unitarity method} to the phase space integrals, allowing integration-by-parts (IBP) 
identities to be utilized analogously to loop integrals. The phase space integration measure 
specific to fragmentation reads~\cite{Bodwin:2012xc}:
\begin{eqnarray}
d\Phi_n &=& 
{8\pi m \over S_n} \delta(k^+-P^+-\sum_{i=1}^n k_i^+) \prod_{i=1}^n \frac{d^{D}k_{i}}{(2\pi)^{D-1}} \delta_+(k_i^2),
\label{phase:space:measure}
\end{eqnarray}
where $k_i$ is the momentum of the $i$-th spectator parton ($g/q/\bar{q}$) 
crossing the cut, and $S_n$ is the statistical factor for identical particles.
The delta functions in \eqref{phase:space:measure} can be eliminated
via the identity $2\pi i\, \delta(x)=\frac{1}{x-i\epsilon}-\frac{1}{x+i\epsilon}$.
Note that the derivation of the IBP identities is insensitive to the 
$i\epsilon$ prescription.
  
Directly applying the IBP method to Eq.~\eqref{phase:space:measure} yields unregulated light-cone singularities that 
cannot be regularized by dimensional regularization alone.
These light-cone singularities are a well-known obstacle in higher-order calculations of quarkonium fragmentation functions~\cite{Feng:2018ulg,Feng:2021uct,Feng:2021qjm,Zhang:2018mlo,Zhang:2020atv} 
(also referred to as rapidity divergences in Ref.~\cite{Zhang:2018mlo}).
To overcome this, we take the reference momentum $ n^\mu $ in the eikonal propagator off the light cone by setting $ r \equiv n^2 \neq 0 $ . This strategy successfully regulates the light-cone singularities and allows for standard IBP reduction~\footnote{Alternatively, the authors of Refs.~\cite{Zhang:2018mlo,Zhang:2020atv} 
circumvent the light-cone singularities by introducing a small gluon mass 
$m_g$ in the phase space integration, taking the limit $m_g \to 0$ after the IBP reduction.}. 
Using this approach, we identify 118 integral families and 1117 master integrals for the double-real corrections, 
as well as 29 families and 258 master integrals for the real-virtual corrections.

The central task is to compute these MIs with high precision in the $r\to 0^+$ limit,
which we achieve using the \textsf{AMFlow} package~\cite{Liu:2017jxz,Liu:2020kpc,Liu:2022mfb,Liu:2022chg}. 
The strategy based on differential equations and the numerical fitting procedure is 
detailed in Appendix~\ref{Appendix:A}.

As a cross-check, we evaluate the double-real and real-virtual corrections using sector decomposition~\cite{Binoth:2000ps,Binoth:2003ak}, as implemented in \textsf{FIESTA}~\cite{Smirnov:2013eza}. 
By bypassing the IBP reduction, this approach has been successfully applied to NLO QCD corrections to 
quarkonium fragmentation functions~\cite{Feng:2018ulg,Feng:2021uct,Feng:2021qjm}. 
We find excellent agreement between the two methods across dozens of $z$ values, 
Nevertheless, we emphasize that the differential equation approach adopted in this work 
yields significantly better numerical precision than sector decomposition.

\section{NNLO fragmentation functions and asymptotic behavior}
\label{NNLO:Frag:Func:Asym:Behavior}

Away from the upper endpoint, both the virtual-real and double-real corrections exhibit 
$1/\epsilon^3$ poles for a given $z\in (0,1)$, a characteristic feature of NNLO calculations. 
Fortunately, these leading poles cancel exactly when the two contributions are summed~\footnote{ Drawing on lessons from NLO calculations~\cite{Braaten:2000pc}, we anticipate that both the virtual-real and double-real sectors exhibit a leading endpoint singularity of the form $\frac{1}{\epsilon^4}\delta(1-z)$, which our numerical algorithm fails to isolate. The double-virtual sector is expected to contain analogous leading poles. Interestingly, our explicit calculation reveals that the (two-loop $\times$ tree) and 
(one-loop $\times$ one-loop) contributions in the double-virtual cut diagrams both contain $1/\epsilon^4$ poles, but they exactly cancel each other. Consequently, no leading quartic pole remains in the double-virtual sector. Therefore, these UV-IR mixed 
$1/\epsilon^4 \delta(1-z)$ pole terms are expected to cancel exactly upon summing the contributions from the virtual-real and double-real sectors, should they exist. Furthermore, we analytically reconstruct the coefficients of the nonvanishing $\frac{1}{\epsilon^n}\delta(1-z)$ ($n=3,2,1$) pole terms in the double-virtual sector.}.
We then perform heavy quark mass and field strength renormalization, 
and renormalize the QCD coupling constant in the $\overline{\rm MS}$ scheme.
After applying the two-loop operator renormalization for the 
$g\to c\bar{c}({}^3S_1^{(8)})$ fragmentation function, we obtain UV- and IR-finite NNLO SDCs for any given 
$z$. The DGLAP renormalization procedure for this fragmentation function through NNLO is detailed 
in Appendix~\ref{Appendix:B}.
 
The coefficients functions associated with $\ln^m \mu^2$,  denoted as 
$C^{\lambda}_{m}(z)$ ($m=1,2$) in Eq.~\eqref{NNLO:SDCs:decomposition}, 
have been derived analytically; their explicit forms are provided in Appendix~\ref{Appendix:C}.   
Note that $C^L_2=0$. 
The $\mu$-independent coefficients functions,  
$\widetilde{C}^{\lambda}_0(z)$ and $C^{\lambda}_0(z)$,  
are computed with exceptionally high numerical precision, and their values are tabulated in 
Tables~\ref{tab:trans}--\ref{tab:unpolar} of Appendix~\ref{Appendix:C}.

Unveiling the logarithmic singularities of the NNLO SDCs near the upper endpoint 
is of great theoretical interest. To this end, we perform a double expansion of the
$\mu$-independent NNLO SDCs around $z=1$:  $C^{\lambda}_0(z)=\sum_{m=-1} \sum_{n=0} f_{m,n} (1-z)^m\ln^n (1-z)$ 
(and analogously for $\widetilde{C}^{\lambda}_0(z)$). 
By evaluating these SDCs with exceedingly high numerical precision at multiple values of $z$
in the vicinity of $z=1$, we extract the expansion coefficients
$f_{m,n}$ ($\tilde{f}_{m,n}$) via a least-$\chi^2$ fit.
Subsequently, the PSLQ algorithm~\cite{ferguson1999analysis} is employed to reconstruct their 
analytic forms up to ${\cal O}((1-z)^{15})$ for $\widetilde{C}^{\lambda}_0(z)$ and up to ${\cal O}((1-z)^{9})$ for $C^{\lambda}_0(z)$.
The most singular terms of $C^{T}_0(z)$ and $\widetilde{C}^{T}_0$ in the limit $z\to 1$ read:
\bseq
\bqa
\label{eq:C^T_0:asy:z=1}
C^{T}_0(z) &=& {1\over \pi (1-z)}\left[18\,\ln^3(1-z) + \left({87\over 2}+54\ln 2\right) \,\ln^2(1-z) \right. 
\\
&+& \left. \left( -\frac{3 \pi ^2}{2}-67+36 \ln^2 2+87 \ln 2\right)\,\ln(1-z) + 
{\cal O}(1) \right] +\cdots,  
\nn \\
\widetilde{C}^{T}_0 (z) &=& {1\over \pi (1-z)} \left[-\ln^2(1-z) + (4-6 \ln 2) \,\ln(1-z) + {\cal O}(1)\right]+\cdots.
\label{eq:asy:z=1}
\eqa
\eseq
Applying the Mellin transform to the leading logarithmic term $[\ln^3(1-z)/(1-z)]_+$,
in Eq.~\eqref{eq:C^T_0:asy:z=1} yields a result consistent with the corresponding leading double-logarithmic term, 
$\alpha_s^3 C_A^2/(2\pi)\ln^4 N$, obtained by expanding the resummed exponent in Refs.~\cite{Chung:2024jfk,Chung:2026mii}. 
The knowledge of the subleading logarithms in Eq.~\eqref{eq:C^T_0:asy:z=1} is highly valuable 
for future resummation of threshold logarithms beyond leading logarithmic accuracy.

\begin{figure}[hbt]
\includegraphics[width=0.32\textwidth]{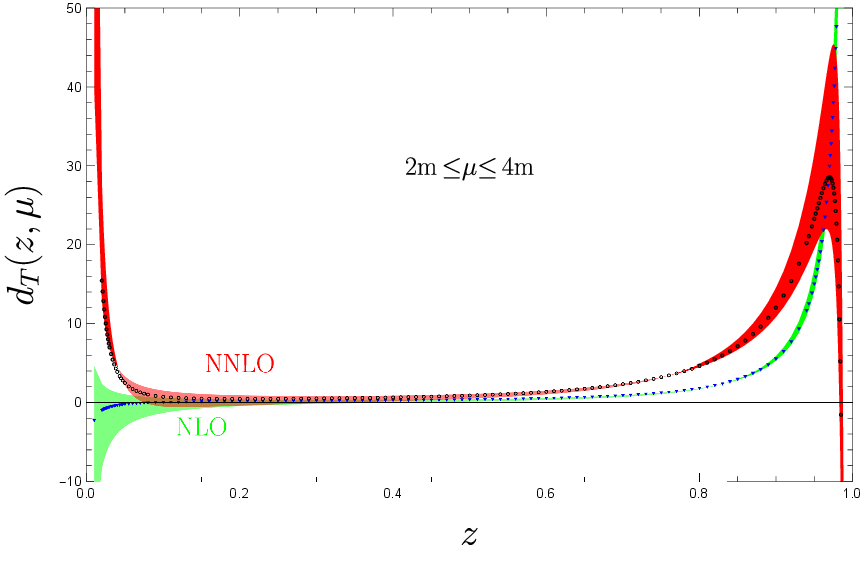}
\includegraphics[width=0.32\textwidth]{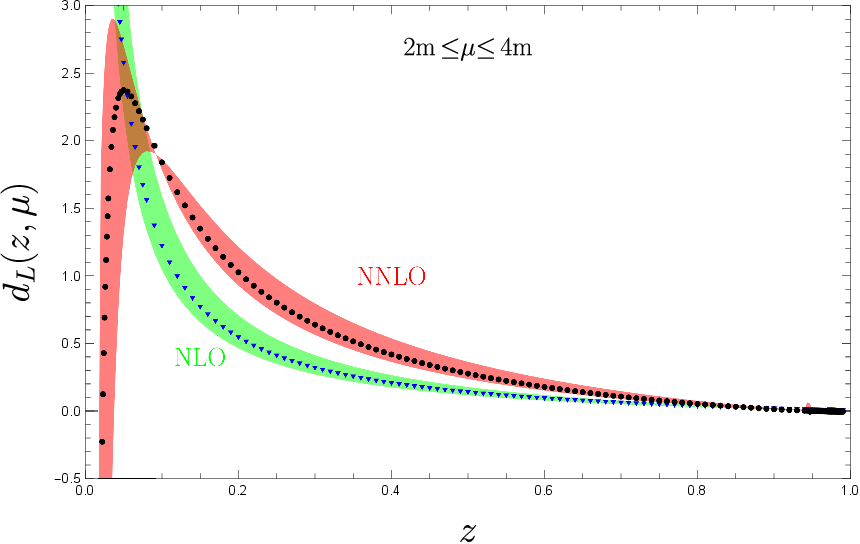}
\includegraphics[width=0.32\textwidth]{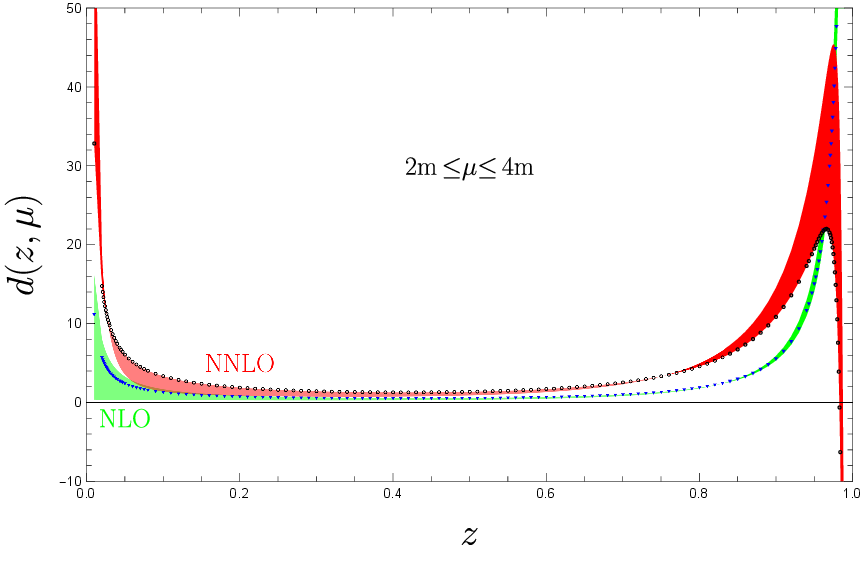}
\caption{SDCs $d_\lambda(z,\mu)$ for $\lambda=T$ (left), $L$ (middle) and 
unpolarized (right) at NLO and NNLO. The NLO and NNLO predictions are defined as 
$d^{\rm NLO}_\lambda(z,\mu)\equiv \alpha^2_s(\mu) \, d^{(1)}_\lambda(z,\mu)$
and $d^{\rm NNLO}_\lambda(z,\mu)\equiv   \alpha_s^2(\mu) \, d^{(1)}_\lambda(z,\mu)+  
\alpha_s^3(\mu) \,d^{(2)}_\lambda(z,\mu)$, respectively.
The shaded bands represent scale uncertainties obtained 
by varying $\mu\in [2m,4m]$.
The strong coupling $\alpha_s$ is evaluated at three-loop accuracy 
using \texttt{RunDec~3}~\cite{Herren:2017osy}.}
\label{NLO:NNLO:3S18FF:plot}
\end{figure}

\begin{figure}[hbt]
\includegraphics[width=0.45\textwidth]{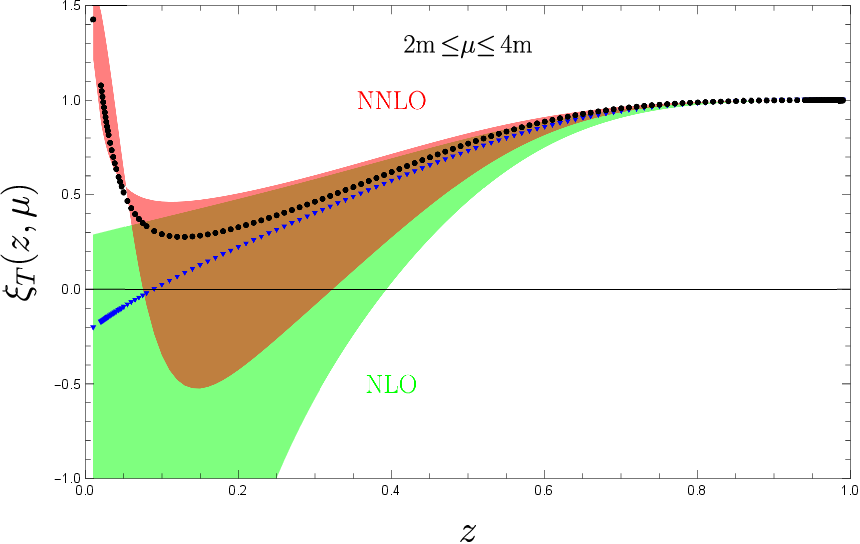}
\includegraphics[width=0.45\textwidth]{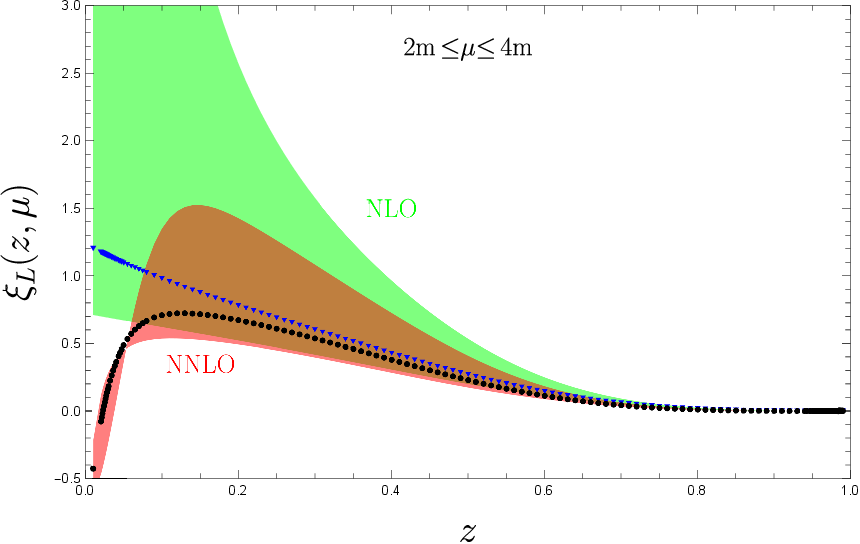}
\caption{Ratios $\xi_\lambda(z,\mu) \equiv {d_\lambda(z,\mu) \over d_T(z,\mu)+d_L(z,\mu)}$ 
for $\lambda=T$ (left) and $L$ (right) at NLO and NNLO. The definitions of the NLO and NNLO predictions follow those
in Fig.~\ref{NLO:NNLO:3S18FF:plot}.
The uncertain bands represent scale variations with $\mu\in [2m, 4m]$.}
\label{NLO:NNLO:polarization:ratio:plot}
\end{figure}

In sharp contrast to the transversely polarized fragmentation function, the NNLO
$\mu$-independent coefficients $C^L_0$ and $\widetilde{C}^L_0$ vanish
as $z\to 1$, consistent with the NLO results. Their asymptotic forms near the upper endpoint are given by
\bseq
\bqa
\label{eq:C^L_0:asy:z=1}
C^{L}_0(z) &=& {1-z\over \pi }\left[-9\ln^2(1-z)- \left(3+18\ln 2\right) \,\ln(1-z)+ {\cal O}(1)\right]+\cdots, 
\\
\widetilde{C}^{L}_0 (z) &=& {1-z\over \pi } \big[\ln(1-z) + {\cal O}(1) \big]+\cdots.
\eqa
\eseq

We further perform asymptotic expansions around  $z=0$ and $z={1\over 2}$, 
determining the coefficients up to ${\cal O}(z^{20)}$ and ${\cal O}((z-{1\over 2})^{20})$,
respectively. Using the PSLQ algorithm, we reconstruct the analytic forms of a sufficient number of 
these coefficients in the former case. 
By combining these local expansions, we achieve a highly accurate global approximation 
across the entire kinematic range $0<z<1$, with relative discrepancies below $10^{-13}$ for $\widetilde{C}^{\lambda}_0(z)$ and $2\times 10^{-6}$ for $C^{\lambda}_0(z)$. This precision is more than adequate for phenomenological applications. 
The numerical values of the coefficients alongside the asymptotic series are provided 
in an ancillary file.

To assess the physical impact of the NNLO corrections, Fig.~\ref{NLO:NNLO:3S18FF:plot} displays 
the SDCs $ d_\lambda(z,\mu) $ as a function of $ z $ for $ \lambda=T,L $ and the unpolarized case 
at NLO and NNLO (Note that the LO result is proportional to $ \delta(1-z) $ and is thus omitted). 
The theoretical uncertainty is estimated by varying the scale $ \mu \in [2m, 4m] $. 
The NNLO corrections to the $g\to c\bar{c}({}^3S_1^{(8)}) $ fragmentation function 
are found to be positive and substantial across most of the $ z $ range.
Notably, the NNLO unpolarized profile develops a distinct peak around $z \approx 0.97$, 
a feature absent at NLO.
Furthermore, Fig.~\ref{NLO:NNLO:polarization:ratio:plot} presents the ratio of the transversely (longitudinally) 
polarized SDC to the unpolarized one at NLO and NNLO.  
Transverse polarization remains dominant at large $z$, 
even after the inclusion of ${\cal O}(\alpha_s^3)$ corrections. 
Additionally, the scale dependence of this polarization ratio 
is significantly reduced from NLO to NNLO.

\section{Summary }

In this work, we have presented the first calculation of the NNLO QCD corrections to the 
$g\to c\bar{c}({}^3S_1^{(8)})$ fragmentation functions within the NRQCD factorization framework 
at the lowest order in the velocity expansion.  
We find that the NNLO corrections are positive and substantial for both transverse and longitudinal polarization states.
To facilitate phenomenological applications, we provide a highly accurate numerical representation of the results 
across the kinematic range $0<z<1$, constructed by matching asymptotic expansions around 
$z=0, 1/2, 1$. Furthermore, the PSLQ algorithm enables the analytic reconstruction of 
logarithmic singularities near the endpoint $z\to 1$, 
providing essential inputs for threshold resummation beyond leading-logarithmic accuracy.

By matching with threshold-resummed formulas in the large-$z$ region, our NNLO predictions provide 
robust inputs for the ${}^3S_1^{[8]}$ gluon fragmentation function. This facilitates a more precise description of 
large-$p_T$ $J/\psi$ ($\psi'$) and $\chi_{cJ}$ production and polarization at hadron colliders, 
marking a pivotal step toward a definitive test of the color-octet mechanism.

 To advance the theoretical understanding of NRQCD factorization, our future research will focus on deriving fully analytic NNLO short-distance coefficients (SDCs) for the $^3S_1^{(8)}$ gluon fragmentation function. This includes determining the various singular distributions at $z=1$, such as $\delta(1-z)$ and the associated plus distributions. Achieving this goal will also yield an important byproduct: the extraction of the two-loop anomalous dimension of the $^3S_1^{(8)}$ NRQCD production operator, which will provide a valuable contribution to the NRQCD framework.

\begin{acknowledgments}
Feynman diagrams in this work are drawn with the aid of {\tt JaxoDraw}~\cite{Binosi:2008ig}.
We thank Long-Bin Chen, Wen Chen, Hee Sok Chung and Xiao Liu for useful discussions. 
The work of F.~F. is supported by the NNSFC under Grant No. 12275353.
The work of Y.~J. is supported in part by the NNSFC under Grant No.~12475090.
The work of W.-L. S. is supported by the NNSFC under Grant No.~12375079.
\end{acknowledgments}

\vspace{0.2 cm}

\appendix

\section{Off-light-cone regularization and differential equations}
\label{Appendix:A}

As detailed in Sec.~\ref{sec:outline:calculation}, we employ the off-light-cone regularization to handle the unregulated light-cone singularities. Following the IBP reduction, the loop integrals are expressed as a linear combination of MIs with their corresponding coefficients. We then derive differential equations (DEs) with respect to $r$,
where $r=n^2$ serves as the regulator in this off-light-cone scheme. For a complete basis of MIs, $\vec{J}$,
the DEs take the form:
\bqa
\frac{\partial }{\partial r}\vec{J}=M(r,\epsilon) \vec{J}.
\label{eq-de}
\eqa
Here, the matrix $M(r,\epsilon)$ is Fuchsian at $r=0$, which implies that 
it possesses at most simple poles at this point. This specific analytic structure guarantees that the 
solution admits a series expansion of the form:
\begin{equation}
\label{eq-asy}
\vec{J}(r,\epsilon) = \sum_k \sum_{n=0}^{\infty} \vec{a}_{k,n}(\epsilon)\,r^{\alpha_k + n},
\end{equation}
where $\alpha_k$ represents a finite set of characteristic exponents determined by the DEs. 
By substituting the asymptotic ansatz from Eq.~\eqref{eq-asy} into the DEs in Eq.~\eqref{eq-de}, 
we can determine the exponents $\alpha_k$ through matching the lowest-order terms. 
Subsequently, the expansion coefficients $\vec{a}$ are systematically computed via recurrence relations. 
To anchor these solutions, the necessary initial conditions at $r=r_0> 0$
are evaluated numerically to high precision using the {\tt AMFlow} package~\cite{Liu:2017jxz,Liu:2022chg,Liu:2022mfb,Liu:2020kpc}.

Furthermore, we implement a robust numerical fitting strategy to reconstruct the full $\epsilon$-dependence of 
the loop integrals. In this procedure, we first assign $\epsilon$ to a set of small rational numbers and derive the corresponding asymptotic expansions (Eq.~\eqref{eq-asy}) for each specific value. We then evaluate the physical limit $r \to 0^+$ by multiplying the asymptotic solutions of the MIs by their respective IBP reduction coefficients. Finally, by solving the resulting system of linear equations, we successfully extract the analytic form of the $\epsilon$-dependent loop integrals in the $r \to 0^+$ limit.

\section{DGLAP renormalization of quarkonium fragmentation function through NNLO }
\label{Appendix:B}

As mentioned in Sec.~\ref{NNLO:Frag:Func:Asym:Behavior}, 
the gluon-to-$c\bar{c}({}^3S_1^{(8)})$ fragmentation function remains UV divergent 
piecewise in $z$ after accounting for charm mass, field strength, and QCD coupling constant renormalizations. 
To remove these residual UV divergences, one must perform the DGLAP operator renormalization up to NNLO.

The renormalized fragmentation function for $i\to H$
is related to its bare counterpart via the identity~\cite{Collins:1981uw}:
\beq
D_{i\to H}(z,\mu^2) = \sum_{j=g,q,\bar{q}} {\cal Z}_{ji}(z,\alpha_s(\mu^2)) \otimes D^{\rm Bare}_{j\to H}(z,\mu^2),
\label{DGLAP:renormalize:FF}
\eeq
where the sum runs over the gluon, light quarks, and antiquarks. The matrix $\mathcal{Z}_{ji}(z,\alpha_s(\mu^2))$ denotes the renormalization functions in the $\overline{\text{MS}}$ scheme, and the symbol $\otimes$ represents the convolution defined by
\begin{equation}
A(z)\otimes B(z) \equiv \int_z^1 \frac{dy}{y} A(y) B(z/y) = B(z) \otimes A(z).
\end{equation}

The renormalization function in Eq.~\eqref{DGLAP:renormalize:FF} admits a perturbative expansion in $\alpha_s$:
\bqa
{\cal Z}_{ji}(z,\alpha_s(\mu^2)) &=& \delta_{ji}\delta(1-z) + \sum_{n=1}^{\infty} \alpha_s^n \sum_{k=1}^n \frac{{\cal Z}_{ji,k}^{(n)}(z)}{\epsilon^k}.
\label{Z:perturbative:expansion}
\eqa

The renormalized fragmentation function $D_{i\to H}(z,\mu^2)$
evolves with the scale according to the DGLAP equation~\cite{Gribov:1972ri,Altarelli:1977zs,Dokshitzer:1977sg}:
\begin{equation}
\frac{d}{d\ln\mu^2} D_{i\to H}(z,\mu^2) = \sum_{j=g,q,\bar{q}} {\cal P}_{ji}(z,\alpha_s(\mu^2)) \otimes D_{j\to H}(z,\mu^2).
\end{equation}
Here, ${\cal P}_{ji}(z,\alpha_s(\mu^2))$ represents the time-like splitting kernel, 
which can be expanded as a power series in $\alpha_s$:
\beq
{\cal P}_{ji}(z,\alpha_s(\mu^2)) =  \sum_{n=1} \alpha_s^n \,{\cal P}^{(n)}_{ji}(z).
\label{P:perturbative:expansion}
\eeq
The leading-order (${\cal O}(\alpha_s)$) splitting kernels, first derived in Ref.~\cite{Altarelli:1977zs}, 
are universal for both time-like and space-like cases. 
However, at next-to-leading order (${\cal O}(\alpha_s^2)$), the time-like splitting kernels 
differ from their space-like counterparts, as detailed in Refs.~\cite{Furmanski:1980cm,Curci:1980uw} 
(see, e.g., Appendix~A of Ref.~\cite{Luo:2019hmp} for a compilation).

Our goal is to extract the renormalization function ${\cal Z}_{ji}(z,\alpha_s(\mu^2))$ up to ${\cal O}(\alpha_s^2)$
from the known time-like splitting kernels. To this end, we note that the splitting kernels are 
related to the renormalization function via
\begin{equation}
{\cal P}_{ji}(z,\alpha_s(\mu^2)) = \sum_{k=g,q,\bar{q}} \frac{d{\cal Z}_{ki}(z,\alpha_s(\mu^2))}{d\ln \mu^2} \otimes {\cal Z}^{-1}_{jk}(z,\alpha_s(\mu^2))).
\label{P:derived:from:Z}
\end{equation}

Substituting Eqs.~\eqref{Z:perturbative:expansion} and \eqref{P:perturbative:expansion} into Eq.~\eqref{P:derived:from:Z} reveals that the ${\cal O}(\alpha_s^n)$ coefficient of the time-like splitting kernel 
is directly related to the single-pole residue of the ${\cal Z}$ function at the same order:
\begin{equation}
{\cal Z}_{ji,1}^{(n)}(z) = -{1\over n}\,{\cal P}^{(n)}_{ji}(z)  \, \quad ({\rm for} \; n \ge 1),
\label{Pn:Zn:relation}
\end{equation}

At NLO, it is sufficient to use
${\cal Z}_{ji,1}^{(1)}(z)=- {\cal P}^{(1)}_{ji}(z)$. 
Combining Eqs.~\eqref{DGLAP:renormalize:FF} and Eq.~\eqref{Pn:Zn:relation}, 
the NLO contribution to the renormalized $i\to H$
fragmentation function can be obtained from the bare one via
\begin{equation}
D_{i\to H}^{(1)}(z)
= D_{i\to H}^{{\rm Bare},(1)}(z) - \sum_{j=g,q,\bar{q}} \frac{1}{\epsilon} \, {\cal P}_{ji}^{(1)}(z) 
\otimes D_{j\to H}^{{\rm Bare},(0)}(z). 
\end{equation}

Performing the DGLAP renormalization up to NNLO requires more than just ${\cal Z}_{ji,1}^{(n)}(z)$
from Eq.~\eqref{Pn:Zn:relation}; the double-pole coefficient $ {\cal Z}_{ji,2}^{(2)}(z) $ in 
Eq.~\eqref{Z:perturbative:expansion} is also needed. Fortunately, 
using the recursive relation~\cite{Becher:2005pd}:
\begin{equation}
 \alpha_s {\partial {\cal Z}_{ji,n+1} \over \partial\alpha_s} = \alpha_s {{\cal Z}_{ki,n} \over \partial \alpha_s} \otimes 
 {\cal Z}_{jk,n} + \beta(\alpha_s)  {\partial {\cal Z}_{ji,n} \over \partial\alpha_s},
\end{equation}
where ${\cal Z}_{ji,n} \equiv \sum_{k=n}^\infty \alpha_s^k {\cal Z}^{(k)}_{ji,n}$,
we can determine ${\cal Z}_{ji,2}^{(2)}(z)$ from the known splitting kernels:
\begin{equation}
{\cal Z}^{(2)}_{ji,2}(z)  = \sum_{k=g,q,\bar{q}} \frac{1}{2} {\cal Z}^{(1)}_{ki,1}(z) \otimes {\cal Z}^{(1)}_{jk,1}(z) - \frac{b_0}{2} {\cal Z}^{(1)}_{ji,1}(z),
\label{Z:second:order:pole:coeff}
\end{equation}
with $b_0 = (11 N_c- 2 n_L)/(12\pi)$.

Combining Eqs.~\eqref{Pn:Zn:relation} and \eqref{Z:second:order:pole:coeff}, and further utilizing Eq.~\eqref{DGLAP:renormalize:FF}, we obtain the NNLO contribution to the renormalized $i\to H$ 
fragmentation function from the bare one via
\begin{eqnarray}
D^{(2)}_{i\to H}(z) & = & D^{{\rm Bare},(2)}_{i\to H}(z) 
- \sum_{j=g,q,\bar{q}} \frac{1}{2\epsilon} \left[ 2{\cal P}_{ji}^{(1)}(z)\otimes D_{j\to H}^{{\rm Bare},(1)}(z) + {\cal P}_{ji}^{(2)}(z)\otimes D_{j\to H}^{{\rm Bare},(0)}(z) \right] 
\nn \\
&& + \sum_{j=g,q,\bar{q}} \sum_{k=g,q,\bar{q}} \frac{1}{2\epsilon^2} \left[  
{\cal P}_{ki}^{(1)}(z) \otimes {\cal P}_{jk}^{(1)}(z) + b_0 {\cal P}_{ji}^{(1)}(z)
\right]\otimes D_{j\to H}^{{\rm Bare},(0)}(z).
\label{DGLAP:renorm:NNLO}
\end{eqnarray}

To renormalize the $g\to c\bar{c}({}^3S_1^{(8)})$ fragmentation function and construct 
the UV counterterms at NNLO, knowledge of the time-like splitting kernels up to ${\cal O}(\alpha_s^2)$
is required. Additionally, we need the bare $g$-to-$c\bar{c}({}^3S_1^{(8)})$ 
 fragmentation function up to ${\cal O}(\alpha_s^2)$, as well as the bare 
 $q$($\bar{q})\to c\bar{c}({}^3S_1^{(8)})$ fragmentation function at the same order, evaluated in 
$d=4-2\epsilon$ spacetime.

In evaluating the UV counterterms in Eq.~\eqref{DGLAP:renorm:NNLO}, 
all convolution integrals are computed analytically and incorporated into the $\mu$-dependent 
coefficient functions $ C_{2,1}^\lambda(z) $ in Eq.~\eqref{NNLO:SDCs:decomposition}.

\section{Results of various NNLO coefficient functions in Eq.~\eqref{NNLO:SDCs:decomposition}}
\label{Appendix:C}

The NNLO coefficient functions associated with the $\ln^2 \mu^2$ term, 
denoted as $C^{\lambda}_{2}(z)$ in Eq.~\eqref{NNLO:SDCs:decomposition}, 
have been derived analytically. We present their explicit expressions below:
\bseq
\begin{eqnarray}
\pi C^{T}_2(z) &=& 
{n_L}\left[\frac{37 z^4-88 z^3+147 z^2-88 z+37}{36(z-1) z}+\frac{1}{3}(z+1) \ln z\right] 
 -\frac{9(z^2-z+1)^2 \ln(1-z)}{(z-1) z} 
 \nn \\ 
&& -\frac{3(11 z^4-30 z^3+93 z^2-30 z+11)}{8(z-1) z} +\frac{9(z^4-4 z^3+3 z^2+1) \ln z}{2(z-1) z},
\\
\pi C^{L}_2(z) &=& 0,
\\
\pi C_2(z)  &=& \pi C^{T}_2(z)+ \pi C^{L}_2(z).
\end{eqnarray}
\eseq

Similarly, the NNLO coefficient functions proportional to 
$\ln \mu^2$, denoted as 
$C^{\lambda}_{1}(z)$ ($\lambda=T,L$) in Eq.~\eqref{NNLO:SDCs:decomposition}, 
are also obtained analytically:
\bseq
\begin{eqnarray}
& & \lefteqn{ \pi C^{T}_1 (z) = n_L \Bigg\{ \left( -\frac{2 z}{3}-\frac{2}{3} \right) \text{Li}_2(1-z) + 
\frac{17 z^2}{108} - \left[ \frac{23 z^2}{9}-\frac{10 z}{3}+\frac{3}{z-1}-\frac{23}{9 z}+\frac{19}{3} \right] \ln(1-z) } 
\nonumber \\
& & + \left[ \frac{13 z^2}{9}-\frac{7 z}{6}+\frac{1}{z-1}-\frac{1}{9 z} - \left( \frac{2 z}{3}+\frac{2}{3} \right) \ln(1-z) - \left( \frac{4 z}{3}+\frac{4}{3} \right) \ln 2 + \frac{11}{6} \right] \ln z 
\nonumber \\
& & - \left[ \frac{37 z^2}{9}-\frac{17 z}{3}+\frac{5}{z-1}-\frac{37}{9 z}+\frac{32}{3} \right] \ln 2 + \frac{13 z}{36} + \frac{1}{z-1} - \frac{89}{108 z} + \left( \frac{z}{3}+\frac{1}{3} \right) \ln^2 z + \frac{47}{36} \Bigg\} 
\nonumber \\
& & + \left[ 18 z^2-18 z-\frac{18}{z}+36 \right] \text{Li}_2(1-z) - \left[ 36 z^2+54 z+\frac{18}{z}+72 \right] \text{Li}_2\left( \frac{z-1}{z} \right) 
\nonumber \\
& & + \left[ 9 z^2+9 z-\frac{9}{z+1}+\frac{9}{z}+18 \right] \text{Li}_2(-z) + 
\left[ 54 z^2+72 \right] \text{Li}_2(z) - \frac{1}{4} \left( 17+36 \pi ^2 \right) z^2 
\nonumber \\
& & + \left[ 27 z^2-27 z+\frac{27}{z-1}-\frac{27}{z}+54 \right] \ln^2(1-z) 
\\
& & - \left[ \frac{27 z^2}{2}+45 z-\frac{27}{4 (z-1)}-\frac{9}{4 (z+1)}+\frac{18}{z}+36 \right] \ln^2 z 
\nn\\
& & + \left[ \frac{3 z^2}{2} + \left( 36 z^2-36 z+\frac{36}{z-1}-\frac{36}{z}+72 \right) \ln 2 - 
\frac{27 z}{2} + \frac{135}{2 (z-1)}-\frac{3}{2 z}+81 \right] \ln(1-z) 
\nonumber \\
& & + \ln z \bigg\{ \frac{15 z^2}{2} + \left[ 27 z^2+63 z-\frac{27}{z-1}+\frac{27}{z}+54 \right] 
\ln(1-z) + \left[ 9 z^2+9 z-\frac{9}{z+1}+\frac{9}{z}+18 \right] \ln(z+1) 
\nonumber \\
& & - \left[ 18 z^2-54 z+\frac{18}{z-1}-\frac{18}{z} \right] \ln 2 + \frac{135 z}{4} - \frac{51}{2 (z-1)} - \frac{15}{2 z} + \frac{45}{4} \bigg\} 
\nonumber \\
& & - \left[ \frac{15 z^2}{2}+\frac{9 z}{2}-\frac{117}{2 (z-1)}-\frac{15}{2 z}-63 \right] \ln 2 + \frac{1}{8} \left( 12 \pi ^2-131 \right) z 
\nonumber \\
& & + \frac{-86-3 \pi ^2}{4 (z-1)} - \frac{3 \pi ^2}{4 (z+1)} + \frac{17+6 \pi ^2}{4 z} + \frac{1}{8} \left( -41-96 \pi ^2 \right)
\nn\\
& & \pi C_{1}^{L}(z) = n_L \left(\frac{2 z^2}{9}-\frac{2 z}{3}-\frac{19}{18 z}+\frac{2 \ln z}{3}+\frac{3}{2}\right)-\frac{3 z^2}{2}+\frac{9 z}{2}-\frac{3}{4 z}+\left(\frac{9}{z}-9\right) \ln (1-z)
 \nonumber\\
&& +\left(-\frac{9}{z}-9\right) \ln (z)-\frac{9}{4}.
\end{eqnarray}
\eseq

For the unpolarized case, the NNLO coefficient function associated with the 
$\ln \mu^2$ term is given by $\pi C_1(z) = \pi C^T_1(z)+ \pi C^L_1(z)$:
\begin{eqnarray}
&& \pi C_1(z)   = \lefteqn{  n_L \Bigg\{ \left( -\frac{2 z}{3}-\frac{2}{3} \right) \text{Li}_2(1-z) + \frac{41 z^2}{108} - 
\left[ \frac{23 z^2}{9}-\frac{10 z}{3}+\frac{3}{z-1}-\frac{23}{9 z}+\frac{19}{3} \right] \ln(1-z) } 
\nonumber \\
& & + \left[ \frac{13 z^2}{9}-\frac{7 z}{6}+\frac{1}{z-1}-\frac{1}{9 z} - \left( \frac{2 z}{3}+\frac{2}{3} \right) \ln(1-z) - \left( \frac{4 z}{3}+\frac{4}{3} \right) \ln 2 + \frac{5}{2} \right] \ln z 
\nonumber \\
& & - \left[ \frac{37 z^2}{9}-\frac{17 z}{3}+\frac{5}{z-1}-\frac{37}{9 z}+\frac{32}{3} \right] \ln 2 - \frac{11 z}{36} + \frac{1}{z-1} - \frac{203}{108 z} + \left( \frac{z}{3}+\frac{1}{3} \right) \ln^2 z + \frac{101}{36} \Bigg\} 
\nonumber \\
& & + \left[ 18 z^2-18 z-\frac{18}{z}+36 \right] \text{Li}_2(1-z) - \left[ 36 z^2+54 z+\frac{18}{z}+72 \right] \text{Li}_2\left( \frac{z-1}{z} \right) 
\nonumber \\
& & + \left[ 9 z^2+9 z-\frac{9}{z+1}+\frac{9}{z}+18 \right] \text{Li}_2(-z) + \left[ 54 z^2+72 \right] \text{Li}_2(z) - \frac{1}{4} \left( 23+36 \pi ^2 \right) z^2 
\nonumber \\
& & + \left[ 27 z^2-27 z+\frac{27}{z-1}-\frac{27}{z}+54 \right] \ln^2(1-z) 
\\
& & - \left[ \frac{27 z^2}{2}+45 z-\frac{27}{4 (z-1)}-\frac{9}{4 (z+1)}+\frac{18}{z}+36 \right] \ln^2 z 
\nonumber \\
& & + \left[ \frac{3 z^2}{2} + \left( 36 z^2-36 z+\frac{36}{z-1}-\frac{36}{z}+72 \right) \ln 2 - 
\frac{27 z}{2} + \frac{135}{2 (z-1)}+\frac{15}{2 z}+72 \right] \ln(1-z) 
\nonumber \\
& & + \ln z \bigg\{ \frac{15 z^2}{2} + \left[ 27 z^2+63 z-\frac{27}{z-1}+\frac{27}{z}+54 \right] \ln(1-z) + \left[ 9 z^2+9 z-\frac{9}{z+1}+\frac{9}{z}+18 \right] \ln(z+1) 
\nonumber \\
& & - \left[ 18 z^2-54 z+\frac{18}{z-1}-\frac{18}{z} \right] \ln 2 + \frac{135 z}{4} - \frac{51}{2 (z-1)} - \frac{33}{2 z} + \frac{9}{4} \bigg\}
\nn\\
& & - \left[ \frac{15 z^2}{2}+\frac{9 z}{2}-\frac{117}{2 (z-1)}-\frac{15}{2 z}-63 \right] \ln 2 + 
\frac{1}{8} \left( 12 \pi ^2-95 \right) z 
\nn\\
& & + \frac{-86-3 \pi ^2}{4 (z-1)} - \frac{3 \pi ^2}{4 (z+1)} + \frac{7+3 \pi ^2}{2 z} + 
\frac{1}{8} \left( -59-96 \pi ^2 \right).
\nn
\end{eqnarray}

\begin{table}[h]
\centering
\setlength{\tabcolsep}{7pt}
\begin{tabular}{ccc||ccc}
\hline\hline
$z$ & $\widetilde{C}_{0}^T(z)$ & $C_{0}^T(z)$ &
$z$ & $\widetilde{C}_{0}^T(z)$ & $C_{0}^T(z)$ 
\\ \hline
 0.050 & -0.67534412 & 792.61593370 & 0.55 & -0.05010144 & -8.46263519 \\ \hline 
 0.10 & 0.47821189 & 132.96591843 & 0.60 & -0.15970760 & -2.37336056 \\ \hline 
 0.15 & 0.55831242 & 23.27550732 & 0.65 & -0.32675554 & 7.47951055 \\ \hline 
 0.20 & 0.49427049 & -6.36889344 & 0.70 & -0.61018429 & 24.06450296 \\ \hline 
 0.25 & 0.40646605 & -15.54360046 & 0.75 & -1.13710545 & 53.19087487 \\ \hline 
 0.30 & 0.32060507 & -18.04322397 & 0.80 & -2.21481550 & 107.27315856 \\ \hline 
 0.35 & 0.24211996 & -18.03505703 & 0.85 & -4.72262284 & 216.47588530 \\ \hline 
 0.40 & 0.17042053 & -16.91188068 & 0.90 & -11.99024201 & 469.98073923 \\ \hline 
 0.45 & 0.10221851 & -15.06256564 & 0.95 & -46.64361062 & 1223.12416714 \\ \hline 
 0.50 & 0.03182655 & 12.39617775 & 0.99 & -621.69705438 & -2320.50242664
 \\ \hline\hline
\end{tabular}
\caption{Numerical results for the $\mu$-independent coefficient functions $\widetilde{C}^{T}_0(z)$ and $C^{T}_0(z)$ corresponding to the transversely polarized fragmentation function, as introduced in Eq.~\eqref{NNLO:SDCs:decomposition}.
\label{tab:trans}}
\end{table}

\begin{table}[h]
\centering
\setlength{\tabcolsep}{7pt}
\begin{tabular}{ccc||ccc}
\hline\hline
$z$ & $\widetilde{C}_{0}^L(z)$ & $C_{0}^L(z)$ &
$z$ & $\widetilde{C}_{0}^L(z)$ & $C_{0}^L (z)$ 
\\ \hline
 0.050 & 1.22145018 & -356.81317646 & 0.55 & -0.16689835 & 9.17018994 \\ \hline 
 0.10 & 0.41980778 & -61.01769397 & 0.60 & -0.16525462 & 7.78044395 \\ \hline 
 0.15 & 0.15716003 & -6.35316499 & 0.65 & -0.16043255 & 6.45294683 \\ \hline 
 0.20 & 0.02698375 & 9.32404515 & 0.70 & -0.15262086 & 5.19193697 \\ \hline 
 0.25 & -0.04875782 & 14.13829232 & 0.75 & -0.14177430 & 3.99764119 \\ \hline 
 0.30 & -0.09622200 & 15.12194231 & 0.80 & -0.12758709 & 2.87021055 \\ \hline 
 0.35 & -0.12685594 & 14.59211625 & 0.85 & -0.10938967 & 1.81400970 \\ \hline 
 0.40 & -0.14652287 & 13.44403076 & 0.90 & -0.08586045 & 0.84660935 \\ \hline 
 0.45 & -0.15854289 & 12.05883543 & 0.95 & -0.05404787 & 0.03243144 \\ \hline 
 0.50 & -0.16491885 & 10.60802236 & 0.99 & -0.01597322 & -0.27021912
 \\ \hline\hline
\end{tabular}
\caption{Numerical values for the $\mu$-independent coefficient functions $\widetilde{C}^{L}_0(z)$ 
and $C^{L}_0(z)$ pertaining to the longitudinally polarized fragmentation function, as introduced in Eq.~\eqref{NNLO:SDCs:decomposition}.
}
\end{table}

\begin{table}[h]
\centering
\setlength{\tabcolsep}{7pt}
\begin{tabular}{ccc||ccc}
\hline\hline
$z$ & $\widetilde{C}_{0}(z)$ & $C_0(z)$ &
$z$ & $\widetilde{C}_{0}(z)$ & $C_0(z)$ 
\\ \hline
 0.050 & 0.54610605 & 435.80275724 & 0.55 & -0.21699980 & 0.70755475 \\ \hline 
 0.10 & 0.89801967 & 71.94822446 & 0.60 & -0.32496222 & 5.40708339 \\ \hline 
 0.15 & 0.71547246 & 16.92234233 & 0.65 & -0.48718809 & 13.93245738 \\ \hline 
 0.20 & 0.52125424 & 2.95515170 & 0.70 & -0.76280515 & 29.25643994 \\ \hline 
 0.25 & 0.35770823 & -1.40530813 & 0.75 & -1.27887975 & 57.18851607 \\ \hline 
 0.30 & 0.22438307 & -2.92128166 & 0.80 & -2.34240259 & 110.14336912 \\ \hline 
 0.35 & 0.11526401 & -3.44294077 & 0.85 & -4.83201252 & 218.28989500 \\ \hline 
 0.40 & 0.02389766 & -3.46784992 & 0.90 & -12.07610246 & 470.82734858 \\ \hline 
 0.45 & -0.05632438 & -3.00373021 & 0.95 & -46.69765849 & 1223.15659858 \\ \hline 
 0.50 & -0.13309230 & -1.78815539 & 0.99 & -621.71302761 & -2320.77264576
\\ \hline\hline
\end{tabular}
\caption{Numerical values for the $\mu$-independent coefficient functions 
$\widetilde{C}_0(z)$ and $C_0(z)$ associated with the unpolarized fragmentation function, 
which are given by the relations $\widetilde{C}_0(z) = \widetilde{C}^T_0(z)+ \widetilde{C}^L_0(z)$ 
and $C_0(z) = C^T_0(z)+ C^L_0(z)$.}
\label{tab:unpolar} 
\end{table}

In this work, the $\mu$-independent NNLO coefficient functions $\widetilde{C}^{\lambda}_0(z)$ and $C^{\lambda}_0(z)$,
are not evaluated analytically. Instead, they are computed with exceptionally high numerical precision, 
and their values are tabulated in Tables~\ref{tab:trans}--\ref{tab:unpolar}. 
To conserve space, the tables present the data rounded to 8 significant digits. 
For applications requiring higher precision, we have provided the complete values with over 100 
significant digits in the accompanying ancillary file.

As shown in Tables~\ref{tab:trans}--\ref{tab:unpolar}, the $\mu$-independent coefficient functions $\widetilde{C}^{T,L}_0(z)$ and $C^{T,L}_0(z)$ diverge as $z\to 0$. In contrast, in the endpoint limit $z\to 1$, both $\widetilde{C}^{T}_0(z)$ and $C^{T}_0(z)$ exhibit severe singularities, whereas $\widetilde{C}^{L}_0(z)$ and $C^{L}_0(z)$ vanish.


\end{document}